\documentclass[prb,a4paper,twocolumn]{revtex4}
\usepackage{amssymb}
\usepackage{t1enc}
\usepackage{amsmath}
\usepackage{amsmath}
\usepackage{epsfig}
\usepackage{amsfonts}
\usepackage{amssymb}

\begin{document}

\title{Conductance of Rashba spin-split systems with ferromagnetic contacts}
\author{Morten H{\o}gsbro Larsen, A. Mathias Lunde, and Karsten Flensberg}
\affiliation{{\O}rsted Laboratory, Niels Bohr Institute fAPG,\\
Universitetsparken 5, 2100 Copenhagen, Denmark. }
\date{\today}

\begin{abstract}
We study theoretically the conductance of heterostructures with ferromagnetic conductors (F)
and a two dimensional electron gas with Rashba spin-orbit interaction (R) using the
Landauer-B\"{u}ttiker formalism. Assuming a one-dimensional model, we first find the
$S$-matrix for the FR interface. This result is then applied to different devices such as a
FRF structure, first suggested by Datta and Das[Appl. Phys. Lett. 56, 665 (1990)]. We find
analytic results for the conductance for the case of collinear magnetization.
\end{abstract}

\maketitle

\section{Introduction}

In 1990 Datta and Das showed that a spin-valve effect arises when a conductor with spin-orbit
coupling of the Rashba type is connected to two ferromagnetic contacts. If furthermore the
spin-orbit interaction can be tuned by an external electric field this effect can maybe be
used as a spin transistor device.\cite{dattadas} The recent years advances in technology of
spin injection into semiconductors has renewed the interest in this type of device. The
Rashba spin-orbit coupling has been measured in a number of materials, e.g. in
heterostructures based on InAs\cite{InAs} or HgTe.\cite{HgTe} Also reports on electric field
control of the Rashba interaction has been reported,\cite{nitta97} but there is still some
discussion about the interpretation of the results.\cite{rowe01}

In this paper we study the conductance of an electron gas with Rashba spin-orbit interaction
(R) sandwiched between ferromagnetic (F) or non-magnetic materials (N) using a convenient
$S$-matrix formalism. Throughout the text we restrict ourselves to a one-dimensional model.
This could be realized by making a point contact structure defined by a set of split gates on
top of the Rashba spin-split 2 dimensional electron gas (2DEG). In higher dimensions the
interference effect due to the Rashba spin-orbit coupling becomes weaker because the phase
shifts depend on the length and the angle of the path between the two contacts.
\section{The model system}

\begin{figure}[tbp]
\centering \epsfig{file=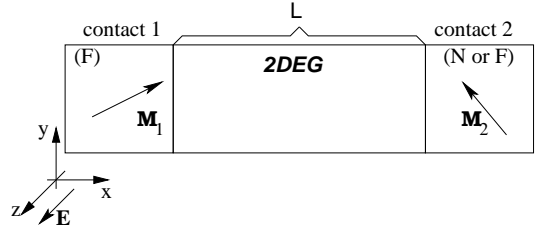,width=7cm} \caption{{\protect\footnotesize Geometry of the
structure studied in this paper.}}
 \label{fig:geometry}
\end{figure}

In this section we specify the Hamiltonian and the eigenstates for the F, R and N segments.
The geometry is as shown in Fig.~\ref{fig:geometry}. The 2DEG has Rashba spin-orbit
scattering corresponding to an electric field in the $z$-direction. The Hamiltonians for the
disorder free regions F,R, and N are, respectively,
\begin{eqnarray}
\widehat{H}_{\mathrm{N}} &=&\frac{\hat{p}^{2}}{2m_{\mathrm{N}}^{{}}}+E_{0}^{N}, \\
\widehat{H}_{\mathrm{F}} &=&\frac{\hat{p}^{2}}{2m_{\mathrm{F}}^{{}}}+\Delta
\frac{\mathbf{M}}{M}
\cdot \widehat{\boldsymbol{\sigma}}+E_{0}^{F}, \\
\widehat{H}_{\mathrm{R}} &=&\frac{\hat{p}^{2}}{2m_{\mathrm{R}}^{{}}}+\frac{\alpha }{ \hbar
}(\widehat{\mathbf{p}}\times \mathbf{E})\cdot \widehat{\boldsymbol{\sigma}}+E_{0}^{R}.
\end{eqnarray}
Here we have defined the spin splitting energy in the ferromagnet $\Delta$ and the
magnetization as $\mathbf{M}=M(\sin \theta \cos \phi ,\sin \theta \sin \phi ,\cos \theta)$,
with $\theta $ and $\phi $ being the usual spherical angles. The magnetizations in two
contacts, $M_1$ and $M_2$, are in general different. Furthermore, $E_{0}^{(F,R,N)}$ is the
band off-sets  and we have used the parabolic band approximation and defined the effective
masses by $m_{\mathrm{F}}^{{}},m_{\mathrm{R}}^{{}}$ and $m_{\mathrm{N}}^{{}}$. The last term
is the Rashba spin orbit interaction where $\alpha $ is the Rashba interaction parameter, and
$\mathbf{E}$ is the field that induces the spin-orbit coupling. The parameter $\alpha |E|,$
which gives the strengths of the coupling, has been argued to be of order $\alpha |E|\sim
10^{-11}$ eVm.\cite{InAs,nitta97}

In order to solve the scattering problem we need to find the velocity operators. We obtain
for the geometry as in Fig.~1
\begin{equation}
\widehat{v}_{\mathrm{R}}^{{}}=\frac{\widehat{p}}{m_{\mathrm{R}}}-\frac{\alpha E
\widehat{\sigma}_y}{\hbar} ,\qquad\widehat{v}_{\mathrm{F}}^{{}}=\frac{\widehat{p}}
{m_{\mathrm{F}}} ,\qquad\widehat{v}_{\mathrm{N}}^{{}}=\frac{\widehat{p}}{m_{\mathrm{N}}}.
\end{equation}
In R the velocity is modified due to the presence of the spin-orbit term.
The velocity
operator is derived for example by noting that $\widehat{v}=i[H,x]/\hbar$
or by using the Hamilton equation $\widehat{v}=\partial_{%
\widehat{p}}^{{}}\widehat{H}$.

Next we find the eigenstates of the Hamiltonians (1)-(3) with energy $\varepsilon $
\begin{eqnarray}
\psi _{\mathrm{F}\sigma }^{\pm }(x) &=&\phi _{\mathrm{F}\sigma }^{\pm }(x)|\sigma \rangle
,\quad \phi _{\mathrm{F}\sigma }^{\pm }(x)=\sqrt{\frac{
m_{\mathrm{F}^{{}}}}{\hbar k_{\sigma }}}e^{\pm ik_{\sigma }x} \\
\psi _{\mathrm{R}s}^{\pm }(x) &=&\phi _{\mathrm{R}s}^{\pm }(x)|s\rangle
,\quad \phi _{\mathrm{R}s}^{\pm }(x)=\frac{1}{\sqrt{v_{\mathrm{R}}}}e^{\pm
ik_{s}x} \\
\psi _{\mathrm{N}\mu }^{\pm }(x) &=&\phi _{\mathrm{N}\mu }^{\pm }(x)|\mu \rangle
,\quad \phi _{\mathrm{N}\mu }^{\pm }(x)=\sqrt{\frac{m}{\hbar k}}e^{\pm ikx}
\end{eqnarray}
where the spinor eigenstates are given by
\begin{eqnarray}
\{|\sigma \rangle ,|\bar{\sigma}\rangle \} &=&\frac{1}{\sqrt{2(1\pm \cos \theta )}}
\begin{pmatrix}
\cos \theta \pm 1 \\
\sin \theta e^{i\phi }
\end{pmatrix}
, \\ \{|s\rangle ,|\bar{s}\rangle \} &=&\frac{1}{\sqrt{2}}
\begin{pmatrix}
\pm i \\1
\end{pmatrix},\\
\{|\mu \rangle ,|\bar{\mu}\rangle \} &=&\left\{
\begin{pmatrix}
1 \\0
\end{pmatrix},
\begin{pmatrix}
0 \\1
\end{pmatrix}
\right\},
\end{eqnarray}
and the relations between the wavenumbers and the eigenenergy $\varepsilon $
are
\begin{eqnarray}
\varepsilon &=&\frac{\hbar ^{2}k_{\sigma }^{2}}{2m_{\mathrm{F}}^{{}}}+\sigma
\Delta +E_{0}^{F},  \label{epsF} \\
\varepsilon &=&\frac{\hbar ^{2}}{2m_{\mathrm{R}}^{{}}}\left[
k_{s}^{{}}(k_{s}^{{}}+2sk_{\mathrm{R}}^{{}})\right] +E_{0}^{R},  \label{epsR}
\\
\varepsilon &=&\frac{\hbar ^{2}k^{2}}{2m_{\mathrm{N}}^{{}}}+E_{0}^{N}.
\label{epsN}
\end{eqnarray}
Here $s$ and $\sigma$ is $\pm 1$ and we have defined
\begin{equation}
k_{\mathrm{R}}^{{}}=\frac{\alpha Em_{\mathrm{R}}^{{}}}{\hbar^{2}}.
\end{equation}
The velocity in the Rashba material $v_{\mathrm{R}}^{{}}$ is
\begin{equation}
v_{\mathrm{R}}^{{}}=\frac{\hbar }{m_{\mathrm{R}}^{{}}}\left( k_{s}^{{}}+sk_{R}^{{}}\right)=
\frac{\hbar }{m_{\mathrm{R}}^{{}}}\left[ \frac{2m_{ \mathrm{R}}^{{}}(\varepsilon
-E_{0}^{R})}{\hbar ^{2}}+k_{\mathrm{R}}^{{2}} \right] ^{1/2},
\end{equation}
where the last equality follows from Eq.~(\ref{epsR}). Note that the velocity is independent
of the Rashba spin direction.

\section{Setting up the $S$-matrix for interface between F and R}

Now we consider the scattering problem related to an interface between an F and an R region.
We start by considering a scattering state for an interface without disorder and with an
electron with spin $\sigma $ incoming from the ferromagnetic side. This we use to prove that
the scattering problem becomes diagonal in the spin degree of freedom. The scattering state
is defined by
\begin{equation}
\psi _{RF,\sigma }(x) =\left\{
\begin{array}{cc}
\phi _{F\sigma }^{+}(x)|\sigma \rangle +r_{\sigma \sigma }^{{}}\phi
_{F\sigma }^{-}(x)|\sigma \rangle , & x\in F \\
\sum_{s}t_{\sigma \sigma }^{{}}\alpha _{s\sigma }^{{}}\phi
_{Rs}^{+}(x)|s\rangle , & x\in R%
\end{array}
\right.
\end{equation}
where we have F in $x<0$ and R in $x>0$.

We solve for the scattering state without including the possibility of being
reflected or transmitted in an $|\bar{\sigma}\rangle $. This is possible
because the spin states in R carry the same group velocity and any linear
combination of $|s\rangle $ and $|\bar{s}\rangle $ is thus an eigenstate to
the velocity operator. Therefore we can choose the coefficients $\alpha
_{s\sigma }$ such that the transmitted state at the interface, $x=0$, is
equal to the state $|\sigma \rangle $
\begin{equation}
\sum \alpha _{s\sigma }|s\rangle =|\sigma \rangle \quad \Rightarrow \quad
\alpha _{s\sigma }=\langle s|\sigma \rangle \equiv U_{s\sigma }.
\end{equation}
The continuity conditions at the interface are that the wavefunction and the probability
currents must be continuous
\begin{equation}
\psi _{RF}(0^{-}) =\psi _{RF}(0^{+}),\,\,\,\, \widehat{v}_{\mathrm{F}}^{{}}\psi _{RF}(0^{-})
=\widehat{v}_{\mathrm{R}}^{{}}\psi _{RF}(0^{+}).
\end{equation}
These 4 equations reduce to only 2 linearly independent equations because, as mentioned
above, the $\psi _{RF}$ is eigenstate of $\widehat{v}_{R}$, and we obtain
\begin{eqnarray}
\frac{1}{\sqrt{v_{\sigma }}}\left( 1+r_{\sigma \sigma }\right) &=&\frac{1}
{\sqrt{v_{\mathrm{R}}^{{}}}}t_{\sigma \sigma }^{{}} \\
\sqrt{v_{\sigma }}(1-r_{\sigma \sigma }) &=&\sqrt{v_{\mathrm{R}}^{{}}}
t_{\sigma \sigma }^{{}}.
\end{eqnarray}
This set of equations corresponds to the scattering between two metals with Fermi velocities
given by $v_{\sigma }^{{}}$ and $v_{\mathrm{R}}^{{}}$, respectively. The are readily solved
\begin{equation}\label{tssrss}
t_{\sigma \sigma }^{{}}=\frac{2}{\sqrt{\frac{v_{\sigma }^{{}}}{v_{\mathrm{R}
}^{{}}}}+\sqrt{\frac{v_{\mathrm{R}}^{{}}}{v_{\sigma }^{{}}}}},\quad r_{\sigma \sigma
}=\frac{\sqrt{\frac{v_{\sigma }^{{}}}{v_{\mathrm{R}}^{{}}}}-
\sqrt{\frac{v_{\mathrm{R}}^{{}}}{v_{\sigma }^{{}}}}}{\sqrt{\frac{v_{\sigma
}^{{}}}{v_{\mathrm{R}}^{{}}}}+\sqrt{\frac{v_{\mathrm{R}}^{{}}}{v_{\sigma }^{{}}}}}.
\end{equation}
Here $t_{\sigma \sigma }^{{}}$ and $r_{\sigma \sigma }^{{}}$ are the transmission and
reflection amplitudes written in the basis defined by the spin eigenstates in F. The
transmission and reflection matrices is thus diagonal in this basis
\begin{equation}
t_{\sigma \sigma ^{\prime }}^{{}}=t_{\sigma \sigma }^{{}}\delta _{\sigma
\sigma ^{\prime }}^{{}},\quad r_{\sigma \sigma ^{\prime }}^{{}}=r_{\sigma
\sigma }^{{}}\delta _{\sigma \sigma ^{\prime }}^{{}}.  \label{tss}
\end{equation}
When we want to combine the transmission through the FR interface with
propagation in R, it is more convenient to write the transmission and
reflection matrices in the basis set defined by spin eigenstates in R. These
we denote by $\mathbf{t}^{R}$ and $\mathbf{r}^{R}$ and they are
\begin{eqnarray}
t_{s^{\prime }s}^{R} &=&\sum_{\sigma }\langle s^{\prime }|\sigma \rangle t_{\sigma \sigma
}^{{}}\langle \sigma |s\rangle \quad \text{or\quad }\mathbf{
t}^{R}=\mathbf{U}\mathbf{t}\mathbf{U^{\dagger }},  \label{tr} \\
r_{s^{\prime }s}^{R} &=&\sum_{\sigma }\langle s^{\prime }|\sigma \rangle r_{\sigma \sigma
}^{{}}\langle \sigma |s\rangle \quad \text{or\quad }\mathbf{
r}^{R}=\mathbf{U}\mathbf{r}\mathbf{U^{\dagger }}.  \label{tr2}
\end{eqnarray}

Next we consider transmission from R to F. For this we set up a scattering state where the
incoming electron from the R side at $x=0$ has a definite spin with respect to the F side. For
the same reasons as above we need only consider one F spin channel at a time. We thus write
\begin{equation}
\psi _{FR,\sigma }(x)=\left\{
\begin{array}{cc}
\sum_{s}\left[ \phi _{Rs}^{-}(x)+r_{\sigma \sigma }^{\prime }\phi
_{Rs}^{+}(x)\right] \alpha _{s\sigma }|s\rangle , & x\in R \\
t_{\sigma \sigma }^{\prime }\phi _{F\sigma }^{-}(x)|\sigma \rangle , & x\in F
\end{array}
\right.
\end{equation}
Solving for $t_{\sigma \sigma }^{\prime }$ $r_{\sigma \sigma }^{\prime }$ in the same way as
above, we obtain the reverse transmission and reflection matrices written in the F spin basis
as
\begin{equation}
\mathbf{t}^{\prime }=\mathbf{t},\quad \mathbf{r}^{\prime }=-\mathbf{r}.
\label{ttmrrm}
\end{equation}
When converted into the basis corresponding to R eigenstates, this becomes
\begin{equation}
\mathbf{t}^{\prime R}=\mathbf{U}\mathbf{t}\mathbf{U}^{\dagger },\quad
\mathbf{r}^{\prime R}=-\mathbf{U}\mathbf{r}\mathbf{U}^{\dagger }.
\label{trp}
\end{equation}
As is evident from Eqs.~(\ref{ttmrrm}) and the fact that $\mathbf{tr=rt}$ the $S$-matrix for
the FR interface
\begin{equation}
\mathbf{S=}\left(
\begin{array}{cc}
\mathbf{r} & \mathbf{t}^{\prime } \\
\mathbf{t} & \mathbf{r}^{\prime }
\end{array}
\right) ,
\end{equation}
is indeed unitary.

The one-dimensional scattering problem is easily generalized to include disorder or interface
scattering if the scattering conserves spin, because the scattering problem would still
separate in two parts, one for each spin direction in the ferromagnet. The total $S$-matrix
is then given by the corresponding problem of scattering between two disordered conductors
with different Fermi velocities. In the following we do not specify the the transmission
coefficient between F and R, and the results are therefore general for any type of interface,
as long as spin flip scattering is not present.

\section{Landauer-B\"{u}ttiker formula}

If the contacts are made of ferromagnetic conductors with a \emph{partial
polarization} then the conductance is according to the Landuer-B\"{u}ttiker
formula given by
\begin{equation}\label{LB}
G=\frac{e^{2}}{h}\sum_{\sigma _{L}\sigma _{R}}|t_{\sigma _{L}^{{}}\sigma
_{R}^{{}}}^{{}}|^{2}=\frac{e^{2}}{h}\mathrm{Tr}\left[ \mathbf{t}^{\dagger }\mathbf{t}\right] ,
\end{equation}
where the $t_{\sigma _{L}^{{}}\sigma _{R}^{{}}}^{{}}$ is the transmission amplitude through
the entire structure written in a basis where the left (rigth) spin state is labeled by
$\sigma _{L(R)}$. This result is of course independent of spin basis set. If, however, we
consider a situation where one or both of the contacts are half metallic, i.e. the
polarization is 100\%, then summation over final and initial states of the transmission
matrix $t_{\sigma _{L}^{{}}\sigma _{R}^{{}}}^{{}}$ must be restricted. This means that the
sum over $\sigma _{L}\sigma _{R}$ in Eq.~(\ref{LB}), must include only populated spin states.

Since we are mainly interested in the the spin dependent transmission we consider the
situation where we can neglect multiple reflections at the interfaces and thus only calculate
the conductance to lowest order in the transmission, i.e.,
\begin{equation}
\mathbf{t}^{(1)}=\mathbf{t}_{2}^{\prime }\mathbf{Lt}_{1},
\end{equation}
where $\mathbf{t}_{1}$ and $\mathbf{t}_{2}$ are the transmission amplitudes for the two
interfaces and $\mathbf{L}$ describes the transmission through the middle region. We take the
middle region to be ballistic but with a Rashba spin orbit coupling and hence
\begin{equation}
\mathbf{L}^{R}=\left(
\begin{array}{cc}
e^{ik_{+}L} & 0 \\
0 & e^{ik_{-}L}
\end{array}
\right) .  \label{L}
\end{equation}

If the coherence is longer than $L$ and if the resonant nature of the transmission is
important one should instead use the formula for $\mathbf{t}$ to all orders in $\mathbf{L}$
\begin{equation}
\mathbf{t}=\mathbf{t}_2^{\prime}\sum_{n=0}^{\infty} (\mathbf{L}%
\mathbf{r}_1^{\prime}\mathbf{L}^{\prime}\mathbf{r}_2^{\prime})^n \mathbf{L}%
\mathbf{t}_1 =\mathbf{t}_2^{\prime}(1-\mathbf{L}\mathbf{r}_1^{\prime}\mathbf{%
L}^{\prime} \mathbf{r}_2^{\prime})^{-1}\mathbf{L}\mathbf{t}_1.
\end{equation}

\subsection{Conductance of a FRN\ structure}

As a first application of our formalism we will calculate the transmission properties and
conductance for a FRN structure. The first-order expression for the transmission matrix is in
the Rashba representation given by
\begin{equation}
\mathbf{t}_{\mathrm{FRN}}^{(1)R}=\mathbf{t}_{2}^{\prime }\mathbf{L}^{R}
\mathbf{U}_{1}\mathbf{t}_{1}\mathbf{U}_{1}^{\dagger }
\end{equation}
where $(\mathbf{t}_{2}^{\prime })_{\sigma\sigma'}^{{}}=\delta_{\sigma \sigma'}^{{}} t_2$ and
$(\mathbf{t}_{1})_{\sigma \sigma ^{\prime }}^{{}}=\delta _{\sigma \sigma ^{\prime
}}^{{}}t_{\sigma \sigma }^{{}}$. Inserting this into the conductance formulae for the case of
a partially polarized ferromagnetic contact
\begin{equation}
G_{\mathrm{FRN}}^{(1)} =\frac{e^{2}}{h}\mathrm{Tr}\left[ \left( \mathbf{t}
_{\mathrm{FRN}}^{(1)R}\right) ^{\dagger }\left( \mathbf{t}_{\mathrm{FRN} }^{(1)R}\right)
\right] =\frac{e^{2}}{h}|t_{2}|^{2}\sum_{\sigma }|t_{\sigma \sigma }|^{2},
\end{equation}
which is independent of all angles. This is also the case for a fully polarized ferromagnetic
contact where the conductance becomes
\begin{equation}
G_{\mathrm{FRN}}^{100\%} =\frac{e^{2}}{h}\sum_{s}|t_{\sigma s}|^{2}=
\frac{e^{2}}{h}|t_{2}|^{2}|t_{\sigma \sigma }|^2.  \notag
\end{equation}
which is of course what we expected because the transmission from F to R was diagonal in F
spin representation and the transmission between R and N is independent of spin direction.
Here we have simply reproduced the result obtained by Ref.\ \onlinecite{mole01}.

The FRN is thus not sensitive to the phase shift induced by the Rashba interaction. The phase
shift acts as a spin polarizer which depends on the length $L$. The way to detect the
polarization is of course to replace the normal metal contact by a ferromagnetic contact,
which was the original suggestion by Datta and Das, \cite{dattadas} which we study in the
following section.

\subsection{Conductance of a FRF$'$\ structure}

As a second application we consider the transmission matrix for the Datta and Das structure,
namely a FRF$^\prime$ device. The transmission to first order in the transmissions is $
\mathbf{t}_{\mathrm{FRF}'}^{R(1)}=\mathbf{t}_{2}^{\prime R}\mathbf{L}^{R} \mathbf{t}_{1}^{R},
$ where $\mathbf{t}_{1}$ and $\mathbf{t}_{2}$ denote the transmission for the two interfaces
and $\mathbf{L}^{\textrm{R}}$ is the transmission matrix for the R system defined in
Eq.~(\ref{L}). Inserting the results from the previous sections, one obtains
\begin{equation}
\mathbf{t}_{\mathrm{FRF}^\prime}^{R(1)}=\mathbf{U}_{2}^{{}}\mathbf{t}_{2}^{\prime}
\mathbf{U}_{2}^{\dagger }\mathbf{L}^{R}\mathbf{U}_{1}^{{}}\mathbf{t}_{1}^{{}}
\mathbf{U}_{1}^{\dagger }.
\end{equation}
If the ferromagnetic contacts are \emph{not} 100 \% spin polarized the conductance is found
by inserting this into Eq.~(\ref{LB}). For the case where the two magnetizations are
parallel, this is explicitly found to be
\begin{align}
G_{\mathrm{FRF}^\prime}^{(1)}(\parallel) &=\frac{e^{2}}{h}\Big[
\frac{1}{2}T_{1}T_{2}+\frac{1}{2}S_{1}S_{2}\Big(
1-2\sin ^{2}\delta   \notag \\
&\quad\qquad+2\sin ^{2}(\delta )\sin ^{2}(\phi )\sin ^{2}(\theta )\Big)\Big] \label{GFRF}
\end{align}
where
\begin{eqnarray}
&&\delta =\frac{L(k_{\bar{s}}-k_{s})}{2},\\\notag &&T_{1}=| t_{1,\sigma \sigma
}|^{2}+|t_{1,\bar{\sigma}\bar{\sigma}}|^{2},\quad S_{1}=|t_{1,\sigma \sigma
}|^{2}-|t_{1,\bar   {\sigma}\bar{\sigma}}| ^{2},\quad
\end{eqnarray}
and similarly for $T_{2}$ and $S_{2}$.

We can now use the result in Eq.~(\ref{GFRF}) to study some special cases. Suppose the two
interfaces are equal and the magnetization of the two magnets are either pointing in the same
direction ($\leftleftarrows )$ i.e. $T_{1}=T_{2}$ and $ S_{1}=S_{2}$, or in opposite
directions ($\rightleftarrows )$ which is equivalent to setting $\Delta _{1}=-\Delta _{2}$ in
Eq.~(\ref{epsF}), i.e. $T_{1}=T_{2}$ and $S_{1}=-S_{2}$, then we get
\begin{align}
G_{\mathrm{FRF}}^{(1)}( \leftleftarrows)&=\frac{e^{2}}{h}\Big[
\frac{1}{2}T^{2}+\frac{1}{2}S^{2}
\Big(1-2\sin ^{2}\delta   \nonumber\\
&\quad\qquad+2\sin ^{2}(\delta )\sin ^{2}(\phi )\sin ^{2}(\theta )\Big)\Big],\label{Gppp}
\end{align}
and $G_{\mathrm{FRF}}^{(1)}(\rightleftarrows )$ follows by the replacement $S^{2}\rightarrow
-S^{2}$ in Eq.~(\ref{Gppp}). The minimum conductance that one can achieve by tuning the
Rashba coupling  (through $\delta $) and/or the magnetization direction, is
\begin{equation}
\left. G_{\mathrm{FRF}}^{(1)}\right| _{\min }=\frac{e^{2}}{h} \frac{1}{2}
(T^{2}-|S|^{2})=\frac{2 e^{2}}{h} |t_{\sigma \sigma }|^{2}|t_{\bar{\sigma}\bar{\sigma}}|^{2}.
\end{equation}

The maximum modulation we can have by changing $\delta $ or the angles of
the magnitization is
\begin{equation}
\left. \Delta G_{\mathrm{FRF}}^{(1)}\right|_{\max
  }=\frac{e^{2}}{h} S^{2}=\frac{e^{2}}{h}  \left(T_{\sigma\sigma}-T_{\bar{\sigma}\bar{\sigma}}\right)^2,
\end{equation}
which is proportional to the difference between the velocities in the
ferromagnet.

The magnetoresponse is given by
\begin{eqnarray}
\Delta r &\equiv &\frac{G_{\mathrm{FRF}}^{(1)}(\leftleftarrows )-G_{\mathrm{
FRF}}^{(1)}(\rightleftarrows )}{G_{\mathrm{FRF}}^{(1)}(\leftleftarrows )+G_{
\mathrm{FRF}}^{(1)}(\rightleftarrows )}  \notag \\
&=&\frac{S^{2}}{T^{2}}\Big(\cos 2\delta +2\sin ^{2}(\delta )\sin ^{2}(\phi
)\sin ^{2}(\theta )\Big)  \label{Dr}
\end{eqnarray}

For the case of two similar half-metallic contacts, i.e. 100\% polarization, the conductance
for parallel case is given by setting $T_{1}=T_{2}=S_{1}=S_{2}=t_{\sigma \sigma }^{2}$,
\begin{eqnarray}
&&G_{\mathrm{FRF}}^{(1)}(\leftleftarrows ,100\%)=\frac{e^{2}}{h}|t_{\sigma \sigma
}|^{2}\nonumber\\
&&\times\Big( 1-\sin ^{2}\delta +\sin ^{2}(\delta )\sin ^{2}(\phi )\sin ^{2}(\theta )\Big),
\label{GFRF100p}
\end{eqnarray}
and the anti-parallel is given by setting $T_{1}=T_{2}=S_{1}=-S_{2}=|t_{\sigma \sigma }|^{2}$,
\begin{equation}
G_{\mathrm{FRF}}^{(1)}(\rightleftarrows ,100\%)=\frac{e^{2}}{h}| t_{\sigma \sigma }|^{2}\sin
^{2}\delta\Big( 1 -\sin ^{2}(\phi )\sin ^{2}(\theta )\Big). \label{GFRF100a}
\end{equation}
For the half-metallic case the maximan modulation is  $t_{\sigma \sigma }^{2} $, and the
magnetoresponse as defined in Eq.~(\ref{Dr}) becomes
\begin{equation}
\Delta r=1-2\sin ^{2}\delta +2\sin ^{2}(\delta )\sin ^{2}(\phi )\sin
^{2}(\theta ).
\end{equation}
This expression shows that the magnetoresponse can be tuned between -1 and 1 depending on the
Rashba coupling and the direction of the magnetization through $\phi $ and $\theta .$

\section{Summary}

We found analytic result for the conductance of an FRF structure with collinear
magnetization, Eq.~(\ref{GFRF}). This formula shows that the minimum conductance is given by
$G_{\min }=(e^2/h)|t_{1\sigma }|^{2}|t_{2\bar{
\sigma}}|^{2}+|t_{1\bar{\sigma}}|^{2}|t_{2\sigma }|^{2}$ , where $|t_{(1,2)\sigma}|^{2}$ and
$|t_{(1,2)\bar{\sigma}}|^{2}$ are the transmission probabilities through the interface (1,2)
for the spin direction $\sigma $ and $\bar{\sigma }$, respectively. In order to have a
substantial valve effect the contacts should therefore by half-metallic or the transmission
coefficients must be strongly spin dependent.


\begin{thebibliography}{9}

\bibitem{dattadas} S.~Datta and B. Das, Appl.\ Phys.\ Lett.\ \textbf{56},
665 (1990)

\bibitem{InAs} T. Matsuyama, R. K\"{u}rsten, C. Mei\b nser, and U. Merkt,
Phys.\ Rev.\ B \textbf{61}, 15588 (2000); Y. Sato, T. Kita, S. Gozu, and S.
Yamada, J. Appl. Phys. \textbf{89}, 8017 (2001).

\bibitem{HgTe} X.C. Zhang, A. Pfeuffer-Jeschke, K. Ortner, V. Hock, H.
Buhrmann, C.R. Becker, G. Landwehr, Phys.\ Rev.\ B \textbf{63}, 245305
(2001).

\bibitem{nitta97} J. Nitta, T. Akazaki, H. Takayanagi, and T. Enoki, Phys.\
Rev.\ Lett.\ \textbf{78}, 1335 (1997).

\bibitem{rowe01} A.C.H. Rowe, J. Nehls, R.A. Strading, and R.S. Ferguson,
Phys.\ Rev.\ B \textbf{63}, R201307 (2001).

\bibitem{mole01} L.W. Molenkamp, G. Schmidt, and G.E. Bauer, Phys.\ Rev.\ B
\textbf{64}, R121202 (2001).
\end{thebibliography}
\end{document}